\documentclass[useAMS,usegraphicx]{mn2e}
\usepackage{amsmath}

\title[\textit{XMM--Newton} long look of MR 2251$-$178]{The puzzling X-ray continuum of the quasar MR~2251$-$178}
\author[E. Nardini et al.]
{E.~Nardini,$^1$\thanks{E-mail: e.nardini@keele.ac.uk} J.~N.~Reeves,$^{1,2}$ D.~Porquet,$^3$ 
V.~Braito,$^4$ N.~Grosso,$^3$ J.~Gofford$^1$ \\
$^1$Astrophysics Group, School of Physical and Geographical Sciences, 
Keele University, Keele, Staffordshire ST5 5BG, UK \\
$^2$Department of Physics, University of Maryland Baltimore County, 
1000 Hilltop Circle, Baltimore, MD 21250, USA \\
$^3$Observatoire Astronomique de Strasbourg, CNRS, UMR 7550, 11 rue de l'Universit\'e, 67000 Strasbourg, France \\
$^4$INAF -- Osservatorio Astronomico di Brera, via E. Bianchi 46, 23807 Merate, Italy}

\begin{document}

\date{Released Xxxx Xxxxx XX}

\pagerange{\pageref{firstpage}--\pageref{lastpage}} \pubyear{2014}

\maketitle

\label{firstpage}

%%%%%%%%%%%%%%%%%%%%%%%%%%%%%%%%%%%%%%%%%%%%%%%%%%%%%%%%%%%%%%%%%%%%%%%
\begin{abstract}
We report on a comprehensive X-ray spectral analysis of the nearby radio-quiet quasar MR~2251$-$178, 
based on the long-look ($\sim 400$ ks) \textit{XMM--Newton} observation carried out in November 2011. 
As the properties of the multiphase warm absorber (thoroughly discussed in a recent, complementary work) 
hint at a steep photoionizing continuum, here we investigate into the nature of the intrinsic X-ray emission 
of MR~2251$-$178 by testing several physical models. The apparent 2--10 keV flatness as well as the 
subtle broadband curvature can be ascribed to partial covering of the X-ray source by a cold, clumpy 
absorption system with column densities ranging from a fraction to several $\times 10^{23}$ cm$^{-2}$. 
As opposed to more complex configurations, only one cloud is required along the line of sight in the presence 
of a soft X-ray excess, possibly arising as Comptonized disc emission in the accretion disc atmosphere. 
On statistical grounds, even reflection with standard efficiency off the surface of the inner disc cannot 
be ruled out, although this tentatively overpredicts the observed $\sim 14$--150 keV emission. It is thus 
possible that each of the examined physical processes is relevant to a certain degree, and hence only a 
combination of high-quality, simultaneous broadband spectral coverage and multi-epoch monitoring of 
X-ray spectral variability could help disentangling the different contributions. Yet, regardless of the model 
adopted, we infer for MR~2251$-$178 a bolometric luminosity of $\sim 5$--$7 \times 10^{45}$ erg s$^{-1}$, 
implying that the central black hole is accreting at $\sim15$--25 per cent of the Eddington limit. 
\end{abstract}


%%%%%%%%%%%%%%%%%%%%%%%%%%%%%%%%%%%%%%%%%%%%%%%%%%%%%%%%%%%%%%%%%
\section{Introduction}
Among the X-ray brightest Active Galactic Nuclei (AGN) in the local Universe due to its 2--10 keV 
luminosity largely exceeding 10$^{44}$ erg s$^{-1}$, MR~2251$-$178 ($z \simeq 0.064$; Canizares, 
McClintock \& Ricker 1978) is a spectacular object in every respect. It was the first quasar to be 
detected and identified through X-ray observations (Cooke et al. 1978; Ricker et al. 1978), as well as 
the first one where the presence was established of \textit{warm} absorption by photoionized gas, 
variable in both ionization state and possibly column density over timescales of less than one year 
(Halpern 1984). The source was later found to experience appreciable changes in the X-ray flux over 
periods of $\sim 10$ days (Pan, Stewart \& Pounds 1990), with a tight correlation between the 
ionization parameter of the absorbing gas\footnote{The ionization parameter is defined as $\xi = 
L_\rmn{ion}/nr^2$, where $n$ is the electron density of the gas and $r$ is its distance from an ionizing 
source with 1--1000 Ry luminosity $L_\rmn{ion}$ (Tarter, Tucker \& Salpeter 1969).} and the continuum 
luminosity (Mineo \& Stewart 1993). Narrow absorption lines with a systematic blueshift of $\sim 300$ 
km s$^{-1}$ have been detected in the ultraviolet (UV) due to Ly$\alpha$, N~\textsc{v} and C~\textsc{iv} 
(Monier et al. 2001). In particular, the C~\textsc{iv} doublet apparently vanished in less than four years 
(Ganguly, Charlton \& Eracleous 2001), implying that the UV absorber is truly local to the AGN, and that 
it is possibly one and the same with the soft X-ray warm absorber. Much deeper insights came with the 
advent of high-resolution X-ray spectroscopy. Early \textit{XMM--Newton} Reflection Grating Spectrometer 
(RGS) observations hinted at a multiphase configuration for the warm absorber, likely consisting of 
two or three distinct components (Kaspi et al. 2004), while the \textit{Chandra} High Energy Transmission 
Grating (HETG) spectrum revealed a highly ionized absorption feature in the iron K band, interpreted 
as the Fe~\textsc{xxvi} Ly$\alpha$ line at the sizable outflow velocity of $v_\rmn{out} \sim 0.04c$ 
(Gibson et al. 2005). The corresponding mass-loss rate was calculated to be at least an order of 
magnitude larger than the accretion rate, unless the covering fraction of the outflow is very small. 

The environment of MR~2251$-$178 is likewise exceptional. The quasar lies in the outskirts of a loose, 
irregular cluster with several tens of galaxies (Phillips 1980), and is surrounded by a huge emission-line 
nebula detected in H$\alpha$ and [O~\textsc{iii}] at optical wavelengths out to a distance of $\sim 100$ 
kpc from the central source (Bergeron et al. 1983; Shopbell, Veilleux \& Bland-Hawthorn 1999). The 
knotty and filamentary appearance of this gaseous envelope is more typical of powerful radio galaxies, 
yet MR~2251$-$178 only shows weak radio emission, whose elongated morphology resembles a 
double-lobed jet-like structure (Macchetto et al. 1990). A similar spatial extent and orientation marks 
the ionization cones recently found through deep [O~\textsc{iii}]$\lambda$5007/H$\beta$ and 
[N~\textsc{ii}]$\lambda$6583/H$\alpha$ flux ratio maps (Kreimeyer \& Veilleux 2013). Overall, the 
quasar radiation field can easily account for the ionization state of the nebula, which is arguably the 
most extended around a radio-quiet source. Its origin, however, is still unclear. Diffuse X-ray emission 
along some directions was preliminary reported by Gibson et al. (2005) based on the smoothed HETG 
zeroth-order image taken in 2002. Unfortunately, we cannot safely corroborate these findings through 
the much deeper 2011 HETG data set, since the artificial broadening of the instrumental Point Spread 
Function due to photon pile-up overrides any faint contribution from the halo.\footnote{The AGN flux 
in the 2011 \textit{Chandra} observation was larger by $\sim 50$ per cent with respect to 2002. While 
the effects of pile-up on the Point Spread Function can be qualitatively simulated, they prevent any 
accurate image reconstruction.} 

Due to its key role as the underlying source of ionizing photons for the ambient gas, from nuclear to 
intergalactic scales, a proper knowledge of the shape and behaviour of the intrinsic X-ray continuum is 
highly desirable, yet most of the effort in the wealth of X-ray analyses performed so far has concentrated 
on improving the characterization of the multi-component warm absorber. Incidentally, all the X-ray spectral 
studies actually agree with describing the absorbed continuum by means of a power law of photon index 
$\Gamma \sim 1.6$, with a high-energy exponential cutoff at $\sim 100$ keV (Orr et al. 2001) and 
a soft excess below $\sim 1$ keV (Kaspi et al. 2004). In a recent paper, we have taken advantage of 
coordinated $\sim 400+400$ ks long \textit{Chandra} HETG and \textit{XMM--Newton} RGS observations 
to bring our grasp on the properties of the intervening ionized gas to unprecedented detail and energy 
resolution (Reeves et al. 2013; hereafter R13). Both campaigns were conducted as large programs in 
late 2011, with just a few weeks of separation from one another, and yielded intriguing information on the 
illuminating continuum itself, which below 2 keV is required to be much steeper ($\Gamma > 2$) than 
typically estimated at hard X-rays. Indeed, the analysis of the \textit{Suzaku} spectrum carried out by 
Gofford et al. (2011) had already suggested that the broadband X-ray emission of MR~2251$-$178 can 
be reproduced equally well through a softer power law with $\Gamma \simeq 2$, provided that an additional 
cold-gas column of $N_\rmn {H} \sim 10^{23}$ cm$^{-2}$ covering a moderate fraction of the source is 
introduced. 

In the wake of these indications, here we present a thorough investigation of the 0.3--10 keV EPIC/pn 
spectrum obtained in the 2011 \textit{XMM--Newton} observation, with the aim of understanding the nature 
of the primary photoionizing continuum and unveiling the high-energy physical processes at work in the 
very central regions of this powerful quasar. This work is organized as follows. In Section~2 we provide 
the basic details about the \textit{XMM--Newton} large observing program on MR~2251$-$178, and describe 
the principal steps of our data reduction. Section~3 is dedicated to the spectral analysis, whose results 
and main implications are discussed in Section~4. Our conclusions are drawn in Section~5. Throughout 
this paper we have assumed $H_0=70$~km~s$^{-1}$~Mpc$^{-1}$, $\Omega_m=0.27$ and 
$\Omega_\Lambda=0.73$, in agreement with the latest values of the concordance cosmological 
parameters (Hinshaw et al. 2013). 

%%%%%%%%%%%%%%%%%%%%%%%%%%%%%%%%%%%%%%%%%%%%%%%%%%%%%%%%%%%%%%%%%
\section{Observations and Data Reduction}

The long-look \textit{XMM--Newton} monitoring of MR~2251$-$178 was performed in separate 
exposures over three consecutive satellite orbits, starting on 2011 November 11, 13, and 15, 
respectively (ObsIDs 0670120201--301--401; PI: J.~Reeves). The corresponding data files were 
processed with the Science Analysis System (\textsc{sas}) v12.0, and were inspected with the standard 
software packages. As the source is quite bright, all the EPIC cameras were operated in small window 
mode, in order to avoid the distortion effects of photon pile-up by virtue of the reduced frame time. 
According to the 10--12 keV light curves, no strong background flaring occurred during any of the 
observations; only a modest rise is seen at both the start and the end of the first and the third ones. 
These periods of weak background activity were rejected adopting a threshold of 0.5 counts s$^{-1}$ 
for the pn, and of 0.35 counts s$^{-1}$ for the two MOS detectors. However, due to an average 0.3--10 
keV count rate of $\sim 4.3$ s$^{-1}$ over a 25$\arcsec$ extraction region, the MOS spectra turned out 
to be still moderately piled-up, and have not been used in this work. 

\begin{figure}
\includegraphics[width=8.5cm]{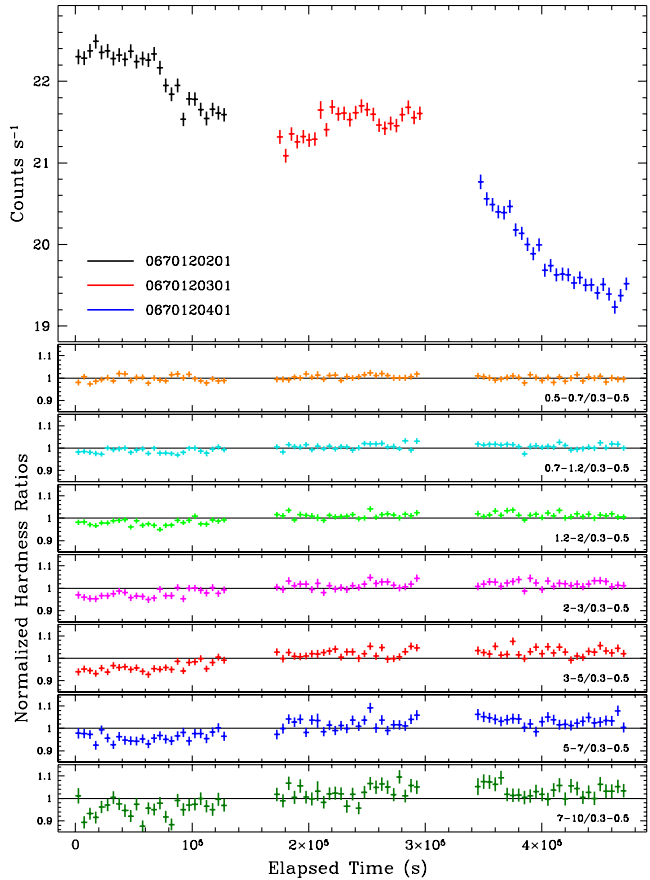}
\caption{Top panel: light curve of MR~2251$-$178 in the 0.3--10 keV energy band 
after background subtraction and dead-time correction, plotted with a time bin of 
5 ks. The peak-to-peak variation in the count rate is 15 per cent. Bottom panels, 
respectively from above: 0.5--0.7, 0.7--1.2, 1.2--2, 2--3, 3--5, 5--7, 7--10 over 
0.3--0.5 keV hardness ratios, normalized to their best-fitting constant values to 
stress the limited extent of any spectral change.} 
\label{lc}
\end{figure}

The EPIC/pn spectra of the source were extracted from a circular region with radius of 36$\arcsec$ 
centred on the target, while the background was evaluated on two identical areas of smaller size 
(31$\arcsec$) at $\sim 3\farcm$5 of distance. The same extraction regions were used for all 
the data sets, in view of their possible merging in the absence of any significant spectral variability 
during the span of the entire campaign. Indeed, Fig.~\ref{lc} shows the background-subtracted 
0.3--10 keV light curve of MR~2251$-$178, along with seven narrow-band hardness ratios with 
respect to the 0.3--0.5 keV energy range. While the overall flux is slowly decreasing, there is no 
evidence of conspicuous changes in the spectral shape. This was also confirmed by the comparison of 
the three X-ray spectra averaged over the separate observations, which were therefore combined 
into a single one. Given the lower pn live time in small window mode (71 per cent), the total effective 
exposure is $\sim 270$ ks. Individual redistribution matrices and ancillary response files at the 
source position were created through the \textsc{sas} tasks \texttt{rmfgen} and \texttt{arfgen}, 
and were subsequently merged as well. The X-ray data were grouped to a minimum of 200 counts 
per energy channel, and the spectral analysis was performed using the \textsc{xspec} v12.8 fitting 
package. All the uncertainties are quoted at the 90 per cent confidence level ($\Delta \chi^2 = 2.71$) 
for the single parameter of interest, while line energies are given in the rest frame, unless otherwise 
stated.

We have also taken into account the target photometry in the six wide-band filters of the Optical 
Monitor (OM), whose data were reprocessed with the \texttt{omichain} pipeline. Only the small 
high-resolution window at the centre of the field of view has been considered. Count rates were 
averaged over the different exposures, with a 10 per cent uncertainty added in quadrature owing 
to possible systematics in the fluxes of the standard stars. An absorption correction was also 
applied, based on the foreground dust reddening maps of Schlegel, Finkbeiner \& Davis (1998) and 
the extinction law of Cardelli, Clayton \& Mathis (1989). With $E(B-V)=0.0390(\pm0.0013)$, averaged 
within a 5$\arcmin$ radius from MR~2251$-$178, and $R_V=A(V)/E(B-V)=3.1$ as per the standard value 
for the diffuse interstellar medium, we obtain $A(V)=0.121(\pm0.004)$ mag, which leads to a flux 
correction ranging from $\sim 1.12$ (V band) to 1.42 (UVW2 band).

%%%%%%%%%%%%%%%%%%%%%%%%%%%%%%%%%%%%%%%%%%%%%%%%%%%%%%%%%%%%%%%%%
\section{Spectral Analysis}

Limited to the X-ray domain, there are still many questions regarding MR~2251$-$178. While further 
progress on the side of the warm absorber above 2 keV will perhaps be feasible only with the new 
technology of X-ray micro-calorimeters onboard the future \textit{Astro--H} mission, at present the most 
critical point involves the origin and intrinsic properties of the continuum. The photon index observed in 
this source is unusually hard compared to the average value of $\Gamma \sim 1.9$ generally found among 
radio-quiet quasars in the same X-ray luminosity range (e.g. Piconcelli et al. 2005; Scott et al. 2011). 
Yet the intensity of the warm absorption lines entails a much softer radiation field below 2 keV ($\Gamma 
\sim 2.5$; R13), denoting either an intrinsically steep continuum (which could be missed if some fraction 
of it were intercepted by cold gas layers still left unmodelled) or an inherent soft excess (e.g. Done et al. 
2012). After a brief overview of the main results from our recent high-resolution analysis, in this Section 
we explore in depth the partial covering conjecture and other possible interpretations for the broadband 
X-ray emission of MR~2251$-$178.

\subsection{Review of RGS and HETG observations}

As discussed in detail in R13, both the \textit{XMM--Newton} RGS and the almost coeval \textit{Chandra} 
HETG spectra reveal the presence of multiple, partially ionized gas components along the line of sight 
to the primary X-ray source in MR~2251$-$178. After a preliminary Gaussian fit of the individual lines 
for parameterization and identification purposes, the absorption spectra were modelled by means of 
self-consistent photoionization grids generated through the \textsc{xstar} code v2.2 (e.g. Kallman \& 
Bautista 2001), and implemented within \textsc{xspec} as multiplicative tables. Three fully-covering 
warm absorption components were required in order to account for the wealth of atomic transitions 
and the wide ionization state of the gas. The generic spectral grid was computed assuming a power-law 
X-ray continuum with $\Gamma=2$. The turbulence velocity was set to $\sigma_\rmn{t} = 100$ km 
s$^{-1}$, based on the narrow profile of most lines; larger values result in substantially worse fits. 
The electron density, instead, is not a highly sensitive parameter. The corresponding tables provide a 
good fit to the mid- to high-ionization lines, due to Fe L-shell ions as well as He- and H-like species 
of lighter elements like C, N, O, Ne, Mg and Si. A separate grid, finely tuned over a limited region of 
the parameter space and involving a steeper illuminating continuum ($\Gamma=2.5$), was introduced 
to better match the intensity of the low-ionization inner-shell lines, including the Fe M-shell Unresolved 
Transition Array (UTA; e.g. Behar, Sako \& Kahn 2001). Interestingly, this component seems to respond 
to long-term flux changes, thus being in photoionization equilibrium. The kinematics and variability 
properties indicate that warm absorption likely occurs beyond the pc scale, i.e. within the inner Narrow 
Line Region. Table~\ref{t1} summarizes the specifics of each grid and the best-fitting values inferred 
from the RGS analysis (as per R13). 

\begin{table}
\caption{Properties of the warm absorption grids and best-fitting parameters to the 2011 RGS 
spectra for a partial covering model. $\Gamma_\rmn{pl}$: spectral index of the illuminating power 
law. $n_e$: electron density in cm$^{-3}$. $\sigma_\rmn{t}$: turbulence velocity in km s$^{-1}$. 
$\mathcal{S}$: sampling points over the 0.1--20 keV energy range. $N_\rmn{H}$: hydrogen column 
density in cm$^{-2}$. $\xi$: ionization parameter in erg cm s$^{-1}$. $\Delta$: grid range. $\delta$: 
grid step. $v_\rmn{out}$: outflow velocity in km s$^{-1}$. } 
\label{t1}
\begin{tabular}{c@{\hspace{30pt}}c@{\hspace{20pt}}c@{\hspace{20pt}}c}
\hline
Component & (a) & (b) & (c) \\
\hline
$\Gamma_\rmn{pl}$ & 2.5 & 2.0 & 2.0 \\
$\log n_e$ & 10 & 10 & 10 \\
$\sigma_\rmn{t}$ & 100 & 100 & 100 \\
$\log \mathcal{S}$ & 5.0 & 4.0 & 4.0 \\
$\Delta \log N_\rmn{H}$ & 20.7--21.7 & 18--24.5 & 18--24.5 \\
$\delta N_\rmn{H}$ & 1$\times$10$^{20}$& 0.5 dex & 0.5 dex \\
$\log N_\rmn{H,best}$ & 21.32 & 21.16 & 21.52 \\
$\Delta \log \xi$ & 0--3 & 0--5 & 0--5 \\
$\delta \xi$ & 0.2 dex & 0.5 dex & 0.5 dex \\
$\log \xi_\rmn{best}$ & 1.27 & 2.02 & 2.78 \\
$v_\rmn{out}$ & 470 & 460 & 0 \\
\hline
\end{tabular}
\end{table}

In the soft X-rays, both the RGS and HETG spectra are also characterized by a complex interplay 
between absorption and emission components, as several narrow absorption lines are found 
superimposed on broad emission features. Among the latter, the strongest and most statistically 
significant ($\Delta \chi^2 \simeq 390$ following its exclusion from the model) is detected around 
the rest-frame energy of 0.56--0.57 keV, as expected for the O~\textsc{vii} triplet. If resolved into 
the resonance, intercombination and forbidden line, the intensity ratios and velocity width suggest 
an origin of such emission consistent with the Broad Line Region (BLR) scale. Conversely, some 
narrower profiles (like those of N~\textsc{vii} Ly$\alpha$ and Ne~\textsc{ix}) might arise at larger 
distances. The identification of all the soft X-ray emission lines and their main properties against 
a dual partial covering description of the underlying continuum (see below) are listed in Table~\ref{t2}. 
All these components were fitted with simple Gaussian profiles, and have been kept frozen in the 
present analysis whenever included in the models.

\begin{table}
\caption{Soft X-ray emission lines in the RGS spectra. $E$: rest energy in eV. $F$: photon flux 
in 10$^{-5}$ cm$^{-2}$ s$^{-1}$. EW: equivalent width in eV. $\sigma_v$: velocity width 
in 10$^3$ km s$^{-1}$.} 
\label{t2}
\begin{tabular}{ccccccc}
\hline
ID & C~\textsc{vi} & N~\textsc{vi} & N~\textsc{vii} & O~\textsc{vii} & O~\textsc{viii} & Ne~\textsc{ix} \\
\hline
$E$ & 362.8 & 418.5 & 498.8 & 564.5 & 654.6 & 905.1 \\
$F$ & 35.8 & 5.9 & 10.1 & 38.4 & 6.5 & 1.2 \\
EW & 2.6 & 0.6 & 1.0 & 7.6 & 1.6 & 0.8 \\
$\sigma_v$ & 4.5 & 1.5 & 0.4 & 4.4 & 1.3 & 0.1 \\
\hline
\end{tabular}
\end{table}

\subsection{EPIC/pn spectral fitting}

\begin{figure}
\includegraphics[width=8.5cm]{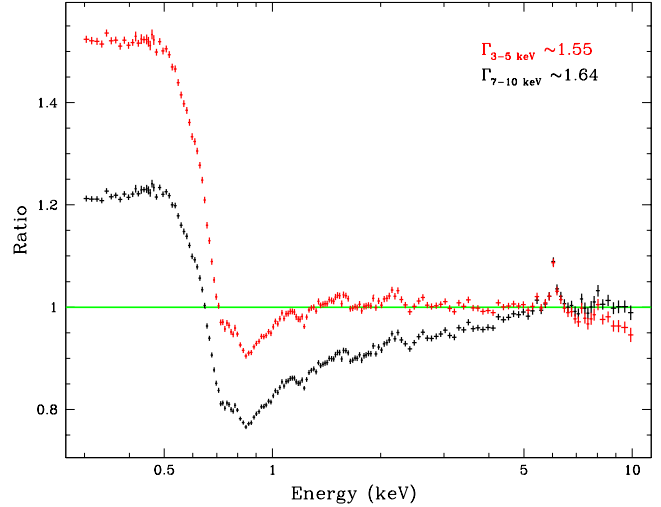}
\caption{Data/model ratios obtained against the best-fitting power laws over the 3--5 (red) 
and 7--10 keV (black) bands, respectively. The X-ray spectral shape of MR~2251$-$178 is 
characterized by an apparent soft excess, a warm absorption trough, a narrow iron emission 
line, and a moderate curvature extending towards the higher energies. The determination of 
the intrinsic continuum is obviously critical to properly model all these observed features. 
(The data were rebinned for clarity).}
\label{rp}
\end{figure}

In agreement with all its previous observations, during the 2011 \textit{XMM--Newton} long look 
the broadband X-ray emission of MR~2251$-$178 exhibits considerable spectral complexity. For 
the sake of illustration, Fig.~\ref{rp} shows the ratio of the 0.3--10 keV EPIC/pn spectrum to a 
power law, assuming two different continuum bands (3--5 and 7--10 keV) as reference. The main 
components, as well as the overall curvature, are clearly brought out. At the same time, it is evident 
that both the depth of the warm absorption trough and the prominence of the contiguous soft excess 
strongly depend on the determination of the intrinsic continuum. To this aim, we have applied widely 
different physical models. In all our fits below, Galactic absorption has been modelled through the 
\texttt{tbabs} cross sections and solar abundances from Wilms, Allen \& McCray (2000), with the 
hydrogen column density frozen to $N_\rmn{H} = 2.4 \times 10^{20}$ cm$^{-2}$ (Kalberla et al. 2005). 

\subsubsection{Partial covering models} 

The sharp detection of tens of absorption lines in the high-resolution spectra amassed unique pieces 
of information on the primary photoionizing continuum, lending weight to a picture where the X-ray 
emission of MR~2251$-$178 is intrinsically much softer than previously thought. The warm absorber, 
however, has a minor effect on the observed slope (especially with its mid- and high-ionization layers), 
and also the global curvature is not fully accounted for. In order to achieve an adequate fit, a partially 
covering cold absorber was employed in the RGS analysis. Besides the superior statistics, this component 
has the advantage of naturally supplying a steeper photon index of $\Gamma \sim 2.3$, very close to the 
one dictated by the low-ionization lines. Reaching out to higher energies, the HETG spectrum accepts 
a second partial covering zone with higher column density to compensate for the residual curvature. 
Both components were reproduced through \textsc{xstar} grids. Due to the abrupt flattening below 
$\sim 0.5$ keV (e.g. Fig.~\ref{rp}), a modest amount (a few $\times$10$^{20}$ cm$^{-2}$) of neutral 
absorption local to the source was also included, possibly associated with the host galaxy rather 
than with the AGN environment. 

\begin{table}
\caption{Best-fitting parameters for the \textsl{pcovpow}, \textsl{discref} and \textsl{pcovref}
physical models (see the text for details) in the 0.3--10 keV spectral range. $\Gamma$: power
law photon index. $K$: power law normalization in 10$^{-2}$ photons keV$^{-1}$ cm$^{-2}$
s$^{-1}$ at 1 keV. $N_\rmn{H}$: column density in cm$^{-2}$. $\xi$: ionization parameter in
erg cm s$^{-1}$. $f$: covering fraction. $\Delta\chi^2$: variation in the fit statistics after
removing a given component from the model and refitting. $Z_\rmn{Fe}$: iron abundance in
solar units. $q$: disc emissivity index for a power-law radial dependence, $\epsilon (r) \propto 
r^{-q}$. $\theta$: disc inclination in degrees. $r_\rmn{in}$: disc inner radius in $r_\rmn{g}$.
$R$: reflection strength. $F_\rmn{obs}$: observed 0.3--10 keV flux in erg s$^{-1}$ cm$^{-2}$.
$L_\rmn{int}$: intrinsic rest-frame 0.3--10 keV luminosity in erg s$^{-1}$.
$\Delta\chi^2_{N_\rmn{H},\xi}$: as above, but keeping the warm absorption $N_\rmn{H}$ and $\xi$ 
fixed to the best-fitting RGS values. (f): frozen parameter.}
\label{t3}
\begin{tabular}{cccc}
\hline
Model & \textsl{pcovpow} & \textsl{discref} & \textsl{pcovref} \\
\hline
\multicolumn{4}{c}{Power Law Continuum} \\
$\Gamma$ & 2.22$\pm$0.03 & 1.77$\pm$0.01 & 1.80$\pm$0.01 \\
$K$ & 4.06$^{+0.27}_{-0.12}$ & 1.00$\pm$0.01 & 1.07$^{+0.03}_{-0.02}$ \\[4pt]
\multicolumn{4}{c}{Partial Covering (1)} \\
$\log N_\rmn{H,1}$ & 22.78$\pm$0.02 & $-$ & 22.67$^{+0.08}_{-0.11}$ \\
$\log \xi_1 $ & 0.96$^{+0.06}_{-0.08}$ & $-$ & 0.0(f) \\
$f_1$ & 0.23$\pm$0.03 & $-$ & 0.09$\pm$0.02 \\[4pt]
\multicolumn{4}{c}{Partial Covering (2)} \\
$\log N_\rmn{H,2}$ & 23.82$\pm$0.02 & $-$ & $-$ \\
$\log \xi_2$ & 0.65$^{+0.35}_{-0.17}$ & $-$ & $-$ \\
$f_2$ & 0.49$^{+0.06}_{-0.04}$ & $-$ & $-$ \\
$\Delta\chi^2$ & 358.9 & $-$ & $-$ \\[4pt]
\multicolumn{4}{c}{Disc Reflection} \\
$\log \xi_\rmn{d}$ & $-$ & 0.80$^{+0.01}_{-0.07}$ & 0.80$^{+0.01}_{-0.14}$ \\
$Z_\rmn{Fe}$ & $-$ & 2.12$\pm$0.25 & 1.75$\pm$0.30 \\
$q$ & $-$ & 3.0$^{+2.7}_{-0.3}$ & 4.4$^{+2.8}_{-1.3}$ \\
$\theta$ & $-$ & 24$^{+3}_{-5}$ & 24$^{+2}_{-1}$ \\
$r_\rmn{in}$ & $-$ & 5.2$^{+4.3}_{-0.9}$ & 9.0$^{+3.2}_{-3.0}$ \\
$R_\rmn{d}$ & $-$ & 1.22$^{+0.23}_{-0.19}$ & 1.08$^{+0.24}_{-0.36}$ \\
$\Delta\chi^2$ & $-$ & 987.1 & 420.7 \\[4pt]
\multicolumn{4}{c}{Remote Reflection} \\
$\log \xi_\rmn{r}$ & $-$ & 2.49$^{+0.02}_{-0.11}$ & 2.17$^{+0.04}_{-0.03}$ \\
$R_\rmn{r}$ & $-$ & 0.08$^{+0.01}_{-0.02}$ & 0.07$\pm$0.01 \\
$\Delta\chi^2$ & $-$ & 127.1 & 108.0 \\[4pt]
\multicolumn{4}{c}{Warm Absorption (a)} \\
$\log N_\rmn{H,a}$ & 21.25$\pm$0.04 & 21.10$\pm$0.05 & 21.19$^{+0.06}_{-0.08}$ \\
$\log \xi_\rmn{a}$ & 1.23$^{+0.11}_{-0.10}$ & 1.24$\pm$0.09 & 1.26$^{+0.06}_{-0.10}$ \\
$\Delta\chi^2$ & 55.5 & 80.2 & 82.6 \\[4pt]
\multicolumn{4}{c}{Warm Absorption (b)} \\
$\log N_\rmn{H,b}$ & 21.49$^{+0.07}_{-0.12}$ & 22.80$^{+0.11}_{-0.47}$ & 21.20$^{+0.13}_{-0.19}$ \\
$\log \xi_\rmn{b}$ & 1.93$^{+0.09}_{-0.04}$ & 3.46$^{+0.02}_{-0.04}$ & 2.26$^{+0.09}_{-0.07}$ \\
$\Delta\chi^2$ & 41.5 & 20.8 & 18.0 \\[4pt]
\multicolumn{4}{c}{Warm Absorption (c)} \\
$\log N_\rmn{H,c}$ & 21.48$^{+0.15}_{-0.11}$ & 21.56$\pm$0.04 & 21.22$^{+0.11}_{-0.20}$ \\
$\log \xi_\rmn{c}$ & 2.79$^{+0.08}_{-0.07}$ & 1.86$^{+0.04}_{-0.03}$ & 1.84$\pm$0.08 \\
$\Delta\chi^2$ & 38.6 & 19.9 & 15.8 \\[4pt]
\multicolumn{4}{c}{General properties} \\
$\log N_\rmn{H}(z)$ & 20.45$\pm$0.03 & $-$ & $-$ \\
$\log F_\rmn{obs}$ & $-$10.19 & $-$10.19 & $-$10.19 \\
$\log L_\rmn{int}$ & 45.32 & 44.87 & 44.88 \\
$\chi^2/\nu$ & 1887.3/1711 & 1899.5/1712 & 1857.7/1710 \\
$\Delta\chi^2_{N_\rmn{H},\xi}$ & 49.4 & 199.3 & 97.3 \\
\hline
\end{tabular}
\end{table}

\begin{figure}
\includegraphics[width=8.5cm]{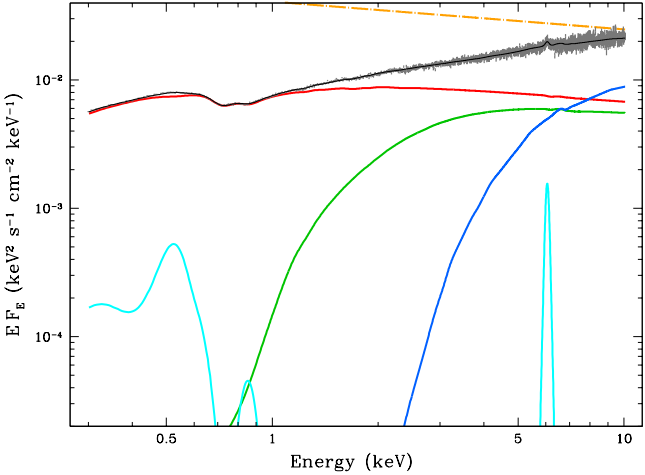}
\caption{Illustration of the \textsl{pcovpow} model and its different spectral components (still convolved 
with the detector's response) against a power law with $\Gamma = 2$. Based on the amount of foreground 
absorption, the intrinsic continuum (shown as a dot-dashed orange line for comparison) is split into three 
parts, respectively transmitted through the warm absorber only (red) and through an additional cold screen 
with either $N_\rmn{H,1} \sim 6 \times 10^{22}$ cm$^{-2}$ (green) or $N_\rmn{H,2} \sim 6.6 \times 10^{23}$ 
cm$^{-2}$ (blue). The blend of soft X-ray emission lines (whose parameters were kept fixed to the RGS-inferred 
ones) is plotted in cyan alongside the narrow Fe K$\alpha$ feature.} 
\label{pc}
\end{figure}

With such a background, we have started our analysis of the 0.3--10 keV spectrum with the straight 
application of this benchmark model, which is dubbed as \textsl{pcovpow} hereafter for ease of discussion. 
Although the warm absorption parameters of Table~\ref{t1} already provide a reasonable fit to the data 
($\chi^2/\nu \sim 1937/1717$), the column densities and ionization states have been allowed to readjust 
to the broadband continuum. Only the outflow velocities were kept fixed, enabling a secure identification 
of each of the three original components. The improvement is significant but not dramatic, leading to 
$\chi^2/\nu \simeq 1887.3/1711$. In most cases the variation of $N_\rmn{H}$ and $\xi$ is minimal (see also 
Section 4), confirming that the warm absorber plays a little role in the 2--10 keV energy range. The 
power-law photon index $\Gamma \simeq 2.22$ lies in between the values inferred from the RGS and HETG, 
and also for the two partial covering components the agreement with the grating analysis appears to be 
excellent. The best-fitting parameters and the general properties of the \textsl{pcovpow} model are 
summarized in Table~\ref{t3}. No discrete detectable lines are connected with the partial covering 
absorbers, as implied by their small degree of ionization, $\log\,(\xi/\rmn{erg\,cm\,s^{-1}}) \la 1$. 
The major impact is clearly on the continuum. With a column density of $N_\rmn{H,1} = 6.0(\pm 0.3) 
\times 10^{22}$ cm$^{-2}$, the first layer is responsible for an appreciable rollover below 2 keV 
(see Fig.~\ref{pc}). This explains why its presence is required in the RGS fit in spite of the narrow 
bandwidth. On the other hand, the evidence for an additional component is now compelling thanks to the 
much larger effective area of EPIC/pn with respect to HETG up to 10 keV. This supplementary screen with 
$N_\rmn{H,2} = 6.6(\pm 0.3) \times 10^{23}$ cm$^{-2}$ and covering fraction $f_2 \sim 0.5$ is in fact 
the main driver of the apparent flatness at hard X-ray energies. Indeed, if a single partial coverer 
were adopted, the data would still prefer an intermediate column of $\sim 2.4 \times 10^{23}$ cm$^{-2}$, 
although the underlying spectral slope would reduce to $\Gamma \sim 1.9$ and, overall, the fit would be 
barely acceptable ($\chi^2_\nu \simeq 1.3$).

Except for the gap of $\sim 6$--7 weeks in the observations, the comparison of EPIC/pn with HETG is 
definitely more informative than that with the simultaneous RGS, given the broader bandpass superposition. 
Focusing on the cold absorbers, both the column densities and the ionization parameters are entirely 
consistent. Only the discrete covering fractions are quite different, but their sum is about 75 per cent at 
both epochs, so that only one fourth of the intrinsic power law is not obscured by any cold gas component. 
As also illustrated in Fig.~\ref{pc}, this means that the absorption-corrected X-ray output of MR~2251$-$178 
exceeds the observed luminosity by a factor of $\sim 3$, once all the putative layers have been disentangled. 
We will touch again on these points in Section 4. Here we note that, in the partial covering scenario depicted 
so far, the apparent soft excess observed below $\sim 0.7$ keV is just a fake feature imprinted on the spectrum 
of MR~2251$-$178 by the combination of warm and cold absorption effects. The other possibility is an actual 
softening of the primary X-ray continuum towards the lower energies. 

\subsubsection{Disc reflection models} 

Soft excesses over the extrapolation of the main power law are found almost ubiquitously among AGN 
with little obscuration (e.g. Porquet et al. 2004), suggesting that they may represent a genuine ingredient 
of the intrinsic X-ray emission. As such, this component has met over the years many alternative 
interpretations, still remaining largely ambiguous. One of the proposed explanations is that the host of soft 
X-ray fluorescent lines from the most abundant elements, produced in the outer layers of the accretion disc 
as a result of the intense X-ray illumination, are blended into a smooth pseudo-continuum due to relativistic 
blurring (Crummy et al. 2006). The tell-tale signature of disc reflection is usually considered the broad, 
skewed component of the Fe K$\alpha$ emission line around 6.4 keV (e.g. Tanaka et al. 1995). This has 
never been reported in MR~2251$-$178, nor seems it to be required here at a simple visual inspection. It 
is well known, however, that this feature is hardly distinguishable from a continuum curvature, possibly 
induced by complex absorption patterns (Turner \& Miller 2009), and this uncertainty works both ways. 

We have therefore concentrated on the 3--10 keV band, attempting to model any iron emission in excess of the 
faint narrow line at $\sim 6.43$ keV (see Table~\ref{t4}). Even though the underlying power law is modified by the 
foreground Galactic column only, which has negligible influence on this range, all the fits are remarkably good, 
with $\chi^2_\nu < 1$. Switching to a broad Gaussian line delivers just a slight improvement ($\Delta\chi^2 \simeq 
-7.5$). We have then maintained the K$\alpha$ profile unresolved, including instead a \texttt{laor} relativistic line 
(Laor 1991), whose shape expresses the motion of the accretion flow within the disc as a function of its inner radius 
(in $r_\rmn{g} = GM_\rmn{BH}/c^2$ units), inclination with respect to the line of sight, and emissivity index 
(complying with a $\epsilon \propto r^{-q}$ radial emissivity profile). As all these parameters cannot be 
constrained at once, we have assumed for the latter two the default values of $\theta=30\degr$ and $q=3$. In 
this case, we achieve a $\Delta\chi^2 \simeq -41$ with the loss of three degrees of freedom, without affecting 
the properties of the narrow core. The inferred equivalent width of the disc line is modest (57$\pm$40 eV), 
and its energy of $\sim 6.54$ keV would imply a moderate disc ionization.

\begin{table}
\caption{Fit to the 3--10 keV band with a power-law continuum plus iron emission 
modelled as: (1) a narrow Gaussian line; (2) a broad Gaussian line, (3) a narrow Gaussian 
line and a \texttt{laor} disc line ($q=3$, $\theta=30\degr$). $E$: line rest energy in keV. 
$\sigma$: line width in keV. EW: line equivalent width in eV. (All the other quantities 
are the same defined earlier in Table~\ref{t3}).} 
\label{t4}
\begin{tabular}{l@{\hspace{30pt}}c@{\hspace{15pt}}c@{\hspace{15pt}}c}
\hline
Model & (1) & (2) & (3) \\
\hline
$\Gamma$ & 1.578$^{+0.006}_{-0.007}$ & 1.580$^{+0.007}_{-0.006}$ & 1.583$^{+0.006}_{-0.007}$ \\
$K$ & 0.840$\pm$0.006 & 0.842$\pm$0.008 & 0.841$\pm$0.009 \\
$E_\rmn{G}$ & 6.429$^{+0.016}_{-0.017}$ & 6.441$^{+0.021}_{-0.026}$ & 6.427$^{+0.019}_{-0.020}$ \\
$\sigma_\rmn{G}$ & 0.01(f) & 0.10$^{+0.04}_{-0.03}$ & 0.01(f) \\
EW$_\rmn{G}$ & 23$\pm$6 & 30$\pm$19 & 17$\pm$7 \\
$E_\rmn{L}$ & $-$  & $-$ & 6.54$^{+0.14}_{-0.11}$ \\
$r_\rmn{in}$ & $-$  & $-$ & $< 3.9$ \\
EW$_\rmn{L}$ & $-$ & $-$ & 57$\pm$40 \\
$\chi^2/\nu$ & 1165.0/1180 & 1157.5/1179 & 1124.4/1177 \\
\hline
\end{tabular}
\end{table}

While this is far from being robust evidence in favour of a broad iron line, it is still worth testing a self-consistent 
treatment of all the reflection effects. Following some previous studies (e.g. Nardini et al. 2011), we have defined 
a model (tagged as \textsl{discref}) consisting of two separate reflection components arising from different locations: 
one is ascribed to the photoionized surface of the accretion disc, and is subject to relativistic blurring (imparted 
through the \texttt{kdblur} convolution kernel); the other is associated with reprocessing material at larger distance, 
and is expected to account for the narrow Fe K$\alpha$ line. The reflection spectra have been modelled with the 
\texttt{xillver} tables of Garc\'ia et al. (2013), and are characterized by iron abundance, ionization state of the gas, 
and photon index of the illuminating power law (assumed to be the same of the intrinsic continuum). These add 
to the key blurring parameters defined above. No partial covering is included, and the soft X-ray emission lines 
have been initially dropped. 

\begin{figure}
\includegraphics[width=8.5cm]{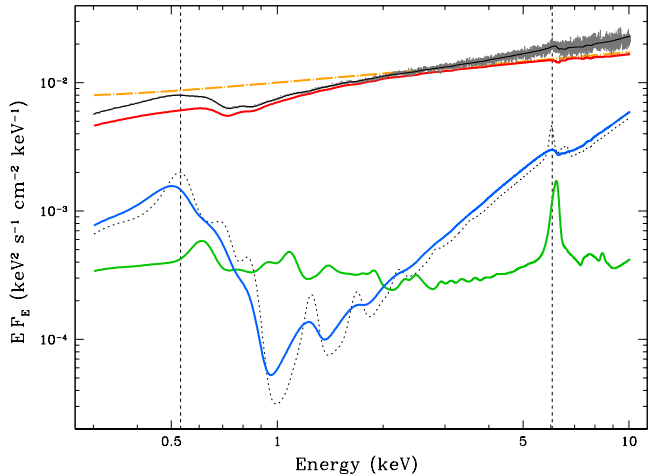}
\caption{Same as Fig.~\ref{pc} for the \textsl{discref} model. The disc component, blurred by the 
relativistic effects, is shown in blue, while the contribution arising from material located much 
farther off the illuminating source is shown in green. The dotted curve portrays the spectral shape 
of disc reflection before the application of any distortions, among which gravitational redshift is 
clearly appreciable. The vertical dashed lines mark the energies of the O~\textsc{vii} soft emission 
feature and of the narrow Fe K$\alpha$ core, respectively.} 
\label{dr}
\end{figure}

The \textsl{discref} model provides a broadly acceptable description of the entire 0.3--10 keV spectrum ($\chi^2/\nu 
\simeq 1899.5/1712$), though slightly worse than \textsl{pcovpow}. All the results are listed in Table~\ref{t3}. 
The disc component is highly significant: once deleted, it is impracticable to recover a sensible fit ($\Delta\chi^2 
\sim 987$). This notwithstanding, as shown in Fig.~\ref{dr}, we are not witnessing a reflection-dominated spectral 
state. While iron abundance is suggested to be around twice solar, the disc inclination is small ($\theta \sim 
25\degr$) and the emissivity index is not so steep to entail a strongly non-isotropic X-ray illumination due to light 
bending (Wilkins \& Fabian 2012). The disc inner radius is consistent with a Schwarzschild black hole. Accepting 
that the disc is not truncated off the innermost stable circular orbit, the dimensionless spin parameter $a^* = 
cJ/GM^2$ (where $J$ and $M$ are the black hole angular momentum and mass) is constrained to be $< 0.65$ 
at the 3$\sigma$ level, whereas the successful application of disc reflection models usually returns rapidly spinning 
black holes (e.g. Walton et al. 2013; Risaliti et al. 2013). This could be due, at least in part, to the differences 
between the \texttt{xillver} and the \texttt{reflionx} (Ross \& Fabian 2005) reflection tables at soft X-ray energies, 
especially for hard illuminations ($\Gamma < 2$; see Garc\'ia et al. 2013 for details). In any case, the reflection 
strength of $R \sim 1.2$ would not be compatible with an extreme gravity regime. This quantity has been computed 
following the definition within the standard \texttt{pexrav} model (Magdziarz \& Zdziarski 1995), where a value 
of $R=1$ is expected in the classical limit for a plane-parallel slab subtending a solid angle of 2$\pi$ at the X-ray 
source. 

\begin{figure}
\includegraphics[width=8.5cm]{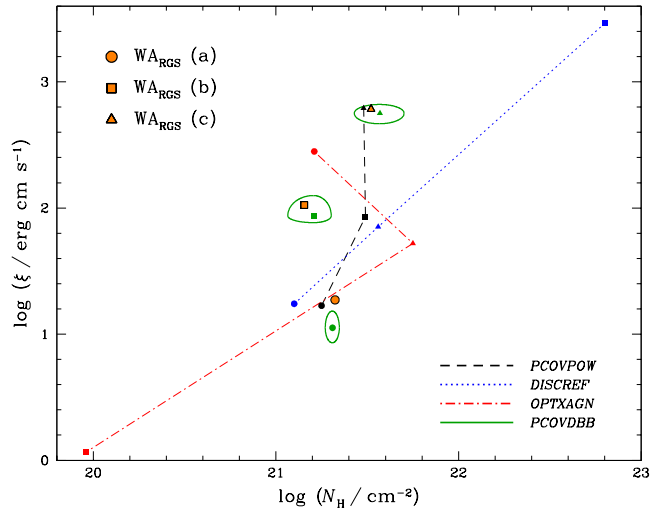}
\caption{Location of the three warm absorbers in the $N_\rmn{H}$--$\xi$ parameter space 
according to the alternative spectral models: \textsl{pcovpow} (black), \textsl{discref} (blue), 
and \textsl{optxagn} (red). Different symbols are used for each component: dots (a), squares (b), 
and triangles (c). The highlighted orange points mark the position of the RGS reference values 
given in Table~\ref{t1}. For simplicity, error ellipses are only shown for the best-fitting \textsl{pcovdbb} 
composite model, showing its excellent agreement with the high-resolution analysis.
We remind that the identification of a given component is based on its outflow velocity.} 
\label{nx}
\end{figure}

It is interesting to point out, however, that the photon index $\Gamma \simeq 1.77$ is still fairly hard, and that 
the ionization state of the two reflection components is somewhat in contrast with the expectations, indicating a 
relatively cold disc with $\log\,(\xi/\rmn{erg\,cm\,s^{-1}}) \simeq 0.8$, against a rather warm distant reflector with 
$\log\,(\xi/\rmn{erg\,cm\,s^{-1}}) \simeq 2.5$. Although the narrow K$\alpha$ core is not formally consistent with an 
origin from neutral gas (Table~\ref{t4}), here it peaks at a rest-frame energy of $\sim 6.6$ keV, which is definitely too 
high. This actually seems to replace the disc line of the phenomenological fit above, as if the smoothing imposed by 
the soft excess were far higher than required by the iron K band. This is a well known issue with broadband reflection 
models. The high ionization of the distant reflector is then likely an artefact to offset the deficiencies of the smeared 
component, including the poor fit to the soft X-ray emission lines, which are resolved by the gratings and have a 
non-relativistic widths.\footnote{Replacing the soft X-ray plus Fe K$\alpha$ emission lines with a single \texttt{xillver} 
table in \textsl{pcovpow} the statistics worsens by $\Delta\chi^2 \simeq 65$. This is likely due to the scattered 
distribution of the reprocessing gas, as suggested by the different velocity widths of the various features (Table~\ref{t2}; 
R13).} To this spectral tweaking significantly contributes the warm absorber, whose properties are strained up to the 
switch between the mid- and high-ionization zones (Fig.~\ref{nx}). Although these two components are individually 
hard to constrain, freezing $N_\rmn{H}$ and $\xi$ at the fiducial values returns a $\Delta\chi^2_{N_\rmn{H},\xi} \sim 
199$ (Table~\ref{t3}). We have therefore attempted to reintroduce the soft X-ray emission-line complex, discarding 
the distant reflector and retaining the narrow Fe K$\alpha$ feature only. The deviations of the warm absorbers are 
indeed reduced, and the fit improves down to $\chi^2/\nu \simeq 1874.4/1711$, yet a clear fine-tuning of the 
relativistic blurring is now involved ($q > 9$, $r_\rmn{in} < 1.3\,r_\rmn{g}$, and $\theta \sim 70\degr$). Hence, we 
did not pursue this speculation any further, sticking to \textsl{discref} in the following for discussion purposes. In 
summary, while we cannot exclude the presence of disc reflection, the related physical conditions are strongly 
model-dependent.

\subsubsection{Comptonization models}

An alternative explanation of the soft excess is Compton up-scattering of the disc photons 
by an electron population with lower temperature and larger optical depth compared to the 
plasma in the hard X-ray corona that gives rise to the main power-law continuum (e.g. 
Haardt \& Maraschi 1993). We have first tested this physical interpretation by applying the 
basic \texttt{compTT} model (Titarchuk 1994), which allowed us to mimic the effects of 
cold/warm Comptonization in a relatively straightforward manner through the Wien temperature 
of the seed photons, and the temperature and optical depth of the scattering electrons. 
More sophisticated models are available (Poutanen \& Svensson 1996; Coppi 1999), but 
the choice of many geometrical and physical variables is in practice largely arbitrary, making 
it not possible to constrain them all simultaneously. 

For consistency, both the soft excess and the hard continuum have been reproduced through 
Comptonized components, with an input photon temperature of 50 eV. The resulting model 
is referred to as \textsl{comp2tt}, and provides an equally effective fit to the data as the 
partial covering and disc reflection ones, with $\chi^2/\nu = 1893.5/1714$. The thermal energy 
of the electrons and the optical depth can be largely degenerate in Comptonization models, 
due to the limited bandpass. We have then frozen the temperature of the hot Comptonizing 
region at $kT_e = 100$ keV, for a corresponding $\tau \sim 0.07$. Quite unexpectedly, it is 
the same optically thin component that becomes sharply dominating below 0.7 keV, entirely 
accounting for the soft excess. Indeed, this is virtually identical to a power law with $\Gamma 
\simeq 2.6$. The optically thick ($\tau \sim 6.5$) counterpart, instead, conforms to the curvature 
of the 2--10 keV spectrum with its anomalously warm temperature of 4 keV. Forcing a more 
realistic value of $kT_e = 0.2$ keV (see e.g. Porquet et al. 2004; Gierli{\'n}ski \& Done 2004), 
the two components are reverted to their expected role, but this gives a much worse fit 
($\Delta \chi^2 > 100$). Another major question raised by this model concerns the nature 
of the X-ray corona and its coupling with the disc, suggesting a complex structure and a high 
degree of inhomogeneity. In principle, though, the temperature of the cold/warm Comptonization 
zone could be the input for further reprocessing. 

\begin{figure}
\includegraphics[width=8.5cm]{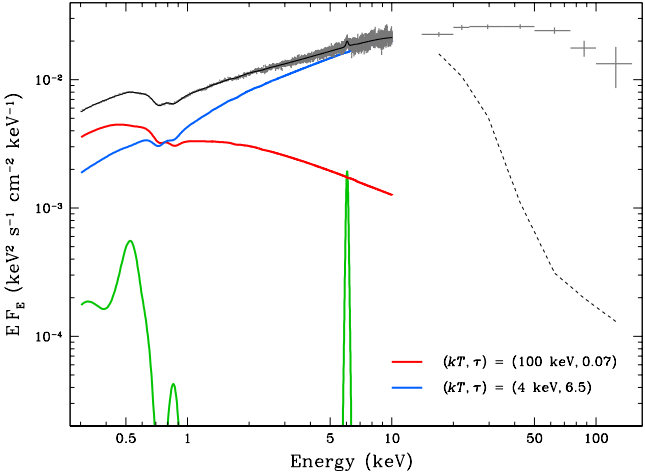}
\caption{Spectral decomposition according to the \textsl{comp2tt} model, with the \textit{inverted} 
soft/hot and hard/warm Comptonized continua plotted in red and blue, respectively (see the text for 
details). The emission-line spectrum is shown in green. The extrapolation towards the higher energies 
reveals the abrupt cutoff that characterizes this configuration. Forcing a standard behaviour (i.e. 
hot power-law tail versus cold soft excess), all the pure Comptonization models are moderately 
unsuccessful at 0.3--10 keV, making this interpretation highly problematic.} 
\label{dc}
\end{figure}

A possible solution to this coronal duality is represented by the \texttt{optxagnf} model 
(Done et al. 2012), in which the three separate components (thermal emission, soft excess 
and power-law continuum) are self-consistently powered by the energy dissipated within 
the accretion flow. A key parameter is the size of the corona ($r_\rmn{cor}$), acting as a 
transitional radius beyond which the emission is characterized by the colour temperature 
corrected blackbody from the outer disc. Below this boundary and down to the innermost 
stable orbit (determined by the black hole spin), the released gravitational energy is 
reprocessed into either the soft excess or the hard power law through Compton 
up-scattering (possibly taking place in the cold/warm disc atmosphere and in the proper 
hot corona, respectively). The power law fraction $f_\rmn{pl}$ indicates the energy balance 
between the two components, while the total luminosity budget depends on the black hole mass 
(estimated to be $2.4 \times 10^8 M_{\sun}$ in MR~2251$-$178; Dunn et al. 2008), its 
spin, and the accretion rate. Due to a substantial degeneracy among the parameters, we 
have frozen $r_\rmn{cor} = 10\,r_\rmn{g}$, and assumed that the source radiates at 10 
per cent of its Eddington luminosity, i.e. $\log\,(L/L_\rmn{Edd}) = -1$. The outer disc 
radius was set to 10$^5\,r_\rmn{g}$. 

Besides the complex continuum encompassing the contributions outlined above, this new model 
(dubbed as \textsl{optxagn}) consists of the usual warm absorption components and of the 
soft X-ray plus Fe K$\alpha$ emission lines. The outcome is a slightly better fit with respect 
to \textsl{comp2tt} ($\chi^2/\nu = 1890.5/1714$), with a power-law photon index of $\Gamma 
\simeq 2.45$ (Table~\ref{t5}), fully consistent with the illumination required by the low-ionization 
warm absorption lines. Contrary to \textsl{discref}, here the central black hole is suggested 
to rotate quite rapidly, with $a^* \sim 0.92$. This is not to be taken at face value, as even a 
minor adjustment of the other quantities (e.g. doubling both $L/L_\rmn{Edd}$ and $r_\rmn{cor}$) 
leaves the goodness of fit virtually unchanged ($\Delta\chi^2 = 3.5$) without implying any spin. 
It is also worth noting that after thawing the Eddington ratio and the coronal radius the fit 
improves by $\Delta\chi^2 = -5.9$ only, delivering $\log\,(L/L_\rmn{Edd}) \simeq -0.5$ and 
$r_\rmn{cor} \sim 5\,r_\rmn{g}$. In any case, the major issue of a twisted spectral decomposition 
is still present, since the temperature ($kT_e \sim 4$ keV) and optical depth ($\tau \sim 7$) of 
the intended soft excess strictly coincide with those of the warm/hard zone in 
\textsl{comp2tt}.\footnote{The Compton temperature of the power-law tail is bound to 100 keV 
within the \texttt{optxagnf} model.} This configuration can be ultimately ruled out considering 
its behaviour above 10 keV, which is clarified by means of \textsl{comp2tt} in Fig.~\ref{dc}. 
Since it is driven by the 2--10 keV curvature, the optically-thick component abruptly fades 
at higher energies, where also its power-law-like counterpart is negligible due to its steepness. 
As a result, the shape of the $\sim 14$--150 keV spectrum obtained by the Burst Alert Telescope 
(BAT) onboard \textit{Swift} in the 70-month all-sky survey (December 2004--September 2010; 
Baumgartner et al. 2013) is completely missed. 

Another severe problem with the Comptonization scenario is the loss of any agreement with 
the gratings on the properties of the warm absorber. Even the pivotal low-$\xi$ component 
turns out to be much more ionized than expected (by a factor of $\sim 15$; Fig.~\ref{nx}). 
If $N_\rmn{H}$ and $\xi$ are not allowed to vary from the RGS measures, the fit is not 
acceptable ($\chi^2_\nu \sim 1.7$ even thawing $L/L_\rmn{Edd}$ and $r_\rmn{cor}$). 
We conclude that pure Comptonization-based models (i.e. not affected by either partial 
covering or disc reflection) are unsuitable to describe the broadband X-ray emission of 
MR~2251$-$178, as it is not possible to reproduce both the overall curvature and the 
double spectral break (i.e. steep--flat--steep) at $E < 1$ keV and $E > 10$ keV (see 
Section 3.3).

\begin{table}
\caption{Best-fitting parameters for the \textsl{optxagn} model. $L/L_\rmn{Edd}$: Eddington 
ratio. $a^*$: dimensionless black hole spin. $r_\rmn{cor}$: radius of the X-ray corona in 
$r_\rmn{g}$. $f_\rmn{pl}$: fraction of the dissipated accretion energy emitted in the hard 
power-law component. (See the text for details).} 
\label{t5}
\begin{tabular}{l@{\hspace{15pt}}c@{\hspace{30pt}}l@{\hspace{15pt}}c}
\hline
$\log\,(L/L_\rmn{Edd})$ & $-1.0$(f) & $\log N_\rmn{H,a}$ & 21.21$^{+0.08}_{-0.09}$ \\
$a^*$ & 0.92$^{+0.03}_{-0.02}$ & $\log \xi_\rmn{a}$ & 2.45$^{+0.08}_{-0.06}$ \\
$r_\rmn{cor}$ & 10(f) & $\log N_\rmn{H,b}$ & 19.96$^{+0.31}_{-0.10}$ \\
$kT_e$ & 3.9$^{+0.3}_{-0.2}$ & $\log \xi_\rmn{b}$ & $< 0.68$ \\
$\tau$ & 6.8$^{+0.5}_{-0.4}$ & $\log N_\rmn{H,c}$ & 21.75$\pm$0.03 \\
$\Gamma$ & 2.45$^{+0.14}_{-0.10}$ & $\log \xi_\rmn{c}$ & 1.72$^{+0.03}_{-0.02}$ \\
$f_\rmn{pl}$ & 0.56$\pm$0.02 & $\log N_\rmn{H}(z)$ & 20.37$^{+0.05}_{-0.03}$ \\
$E_\rmn{G}$ & 6.430$^{+0.017}_{-0.018}$ & $\chi^2/\nu$ & 1890.5/1714 \\
EW$_\rmn{G}$ & 20$\pm$3 & $\Delta\chi^2_{N_\rmn{H},\xi}$ & 1160 \\
\hline
\end{tabular}
\end{table}

\subsubsection{Composite models}

\begin{figure}
\includegraphics[width=8.5cm]{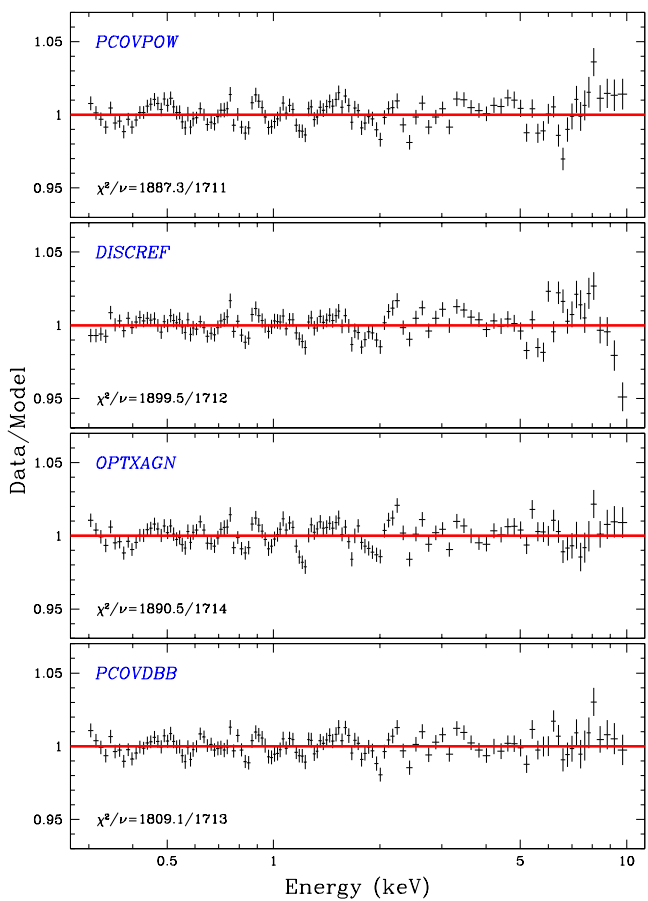}
\caption{Residuals for the different best fits over the 0.3--10 keV energy range, plotted as 
data/model ratios. The spectra have been heavily rebinned in order to emphasize the main 
structures, especially the drop beyond $\sim 9$ keV in the reflection-based model (second 
panel from top).} 
\label{dm}
\end{figure}

While pointing to completely different physical processes, on sheer statistical grounds all the interpretations 
discussed so far are almost equivalent in reproducing the 0.3--10 keV emission of MR~2251$-$178. The fit 
residuals only show limited structures (Fig.~\ref{dm}). The \textsl{discref} and \textsl{optxagn} models have 
several shortcomings, though. Taking a step further, we have then attempted to combine the key components 
of disc reflection and Comptonization with line-of-sight cold absorption, in order to investigate whether these 
processes are simultaneously present. 

We have first introduced a single partial covering layer with ionization parameter fixed to zero in \textsl{discref}, 
assuming the same obscured fraction for the power-law continuum and the blurred reflection component. This 
model is labelled as \textsl{pcovref}, and leads to an improvement of $\Delta\chi^2 \simeq -41.8$ with the loss 
of one degree of freedom.\footnote{The location of the remote reflector with respect to the absorber is not critical. 
For simplicity, this component was left unaffected. Note that, qualitatively, equivalent results are obtained if the 
individual emission lines are used instead.} The full results are listed in Table~\ref{t3}. Since the covering fraction 
is fairly small, just around 0.1, this should be regarded as a refinement of \textsl{discref} rather than as a 
self-standing composite model. The properties of the mid- and high-ionization warm absorbers are still 
misreported, and the pattern in the fit residuals above 9 keV (Fig.~\ref{dm}) is still evident. Nevertheless, 
the cold-gas column of $\sim 5 \times 10^{22}$ cm$^{-2}$, which is very similar to that of the thinner layer 
in \textsl{pcovpow}, indicates that the existence of significant reflection from the disc is mutually exclusive 
with that of a gas screen much thicker than 10$^{23}$ cm$^{-2}$.

\begin{figure}
\includegraphics[width=8.5cm]{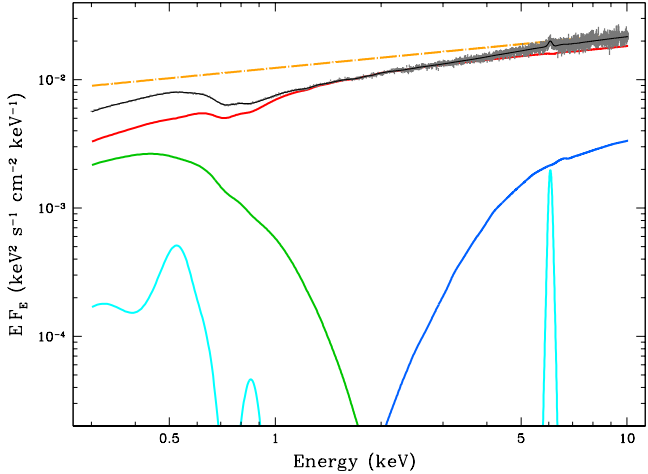}
\caption{Same as Fig.~\ref{pc} for the best-fitting \textsl{pcovdbb} model. In this scenario, the primary 
power-law continuum (dot-dashed orange line) is considerably flatter, and mostly transmitted through 
the warm absorber only (red). Just a fraction of about one sixth (blue) is obscured by an additional 
column of $N_\rmn{H} \sim 3 \times 10^{23}$ cm$^{-2}$. An intrinsic soft excess (green), possibly 
arising as Comptonized disc emission, emerges below 2 keV. The emission-line complex is  plotted in 
cyan.} 
\label{bb}
\end{figure}

On the other hand, it is not straightforward to implement any partial covering within \textsl{optxagn} for normalization 
reasons. We have then opted for the more flexible power law plus soft excess format, modelling the latter component 
with a \texttt{diskbb} blackbody spectrum (e.g. Makishima et al. 1986) for simplicity.\footnote{The nominal lower 
limit for $kT_e$ in \texttt{compTT} is 2 keV, thus the typical temperatures of the soft X-ray excess are not 
sampled. Forcing this restriction, the cold Comptonization component systematically tends to a blackbody-like 
shape in our fits, since $\tau > 30$.} We hence refer to this model as \textsl{pcovdbb}. The fit statistics 
undergoes a remarkable improvement down to $\chi^2/\nu \simeq 1809.1/1713$. Most importantly, this is 
not achieved at the expense of the warm absorption properties, whose conservation is excellent (Fig.~\ref{nx}). 
The results are listed in Table~\ref{t6}. Due to its temperature of $kT_\rmn{in} \sim 0.14$ keV the soft excess 
has no contribution above 2 keV, so that any obscured blackbody emission would be completely obliterated 
by the column density of $N_\rmn{H} \sim 3 \times 10^{23}$ cm$^{-2}$ (Fig.~\ref{bb}). Since this component 
allegedly arises from the very inner regions, we expect the same covering fraction $f \sim 0.17$ of the power 
law to apply. As a consequence, the intrinsic luminosity of the soft excess should be corrected by a factor 
$(1-f)^{-1} \sim 1.2$, accounting for $\sim 20$--25 per cent of the total output of the source at 0.3--2 keV. 
In general, the \textsl{pcovdbb} model appears as the most successful description of the X-ray spectrum 
of MR~2251$-$178.

\begin{table}
\caption{Best-fitting parameters for the \textsl{pcovdbb} model. $kT_\rmn{in}$: temperature at the 
inner disc radius in keV. $L_\rmn{bb}$: observed 0.3--2 keV soft excess flux in erg s$^{-1}$ cm$^{-2}$. 
(Compare with Table~\ref{t3}).} 
\label{t6}
\begin{tabular}{l@{\hspace{15pt}}c@{\hspace{30pt}}l@{\hspace{15pt}}c}
\hline
$\Gamma$ & 1.74$^{+0.02}_{-0.01}$ & $\log N_\rmn{H,b}$ & 21.21$^{+0.09}_{-0.15}$ \\
$K$ & 1.23$\pm$0.03 & $\log \xi_\rmn{b}$ & 1.94$^{+0.16}_{-0.06}$  \\
$kT_\rmn{in}$ & 0.139$\pm$0.003 & $\log N_\rmn{H,c}$ & 21.57$^{+0.13}_{-0.14}$ \\
$\log L_\rmn{bb}$ & $-$11.41$\pm$0.06 & $\log \xi_\rmn{c}$ & 2.75$^{+0.07}_{-0.08}$\\
$\log N_\rmn{H,1}$ & 23.42$^{+0.04}_{-0.03}$ & $\log N_\rmn{H}(z)$ & 20.37$^{+0.04}_{-0.03}$ \\
$f_1$ & 0.17$^{+0.02}_{-0.01}$& $\log L_\rmn{int}$ & 44.96 \\
$\log N_\rmn{H,a}$ & 21.31$\pm$0.04 & $\chi^2/\nu$ & 1809.1/1713 \\
$\log \xi_\rmn{a}$ & 1.05$^{+0.13}_{-0.12}$ & $\Delta\chi^2_{N_\rmn{H},\xi}$ & 26.8 \\
\hline
\end{tabular}
\end{table}

\subsection{Optical to X-ray spectral energy distribution} 

As anticipated for the unrestricted Comptonization models, a viable method to distinguish among the 
competing scenarios is to test the models over a broader spectral range, especially at higher energies, 
where a diverging behaviour is most likely. In principle, the 2009 \textit{Suzaku} spectrum analysed by 
Gofford et al. (2011) could be very instructive in this respect, yet the current models are too intricate for 
that data quality. The same \textit{Suzaku}/PIN detection is quite noisy and provides virtually no constraints 
beyond $\sim 30$ keV. We have therefore retrieved the $\sim 14$--150 keV 70-month \textit{Swift}/BAT 
spectrum, which should be representative of the average intrinsic emission in the very hard X-ray domain 
as it is not affected by the column densities at issue. In terms of a simple power law, the BAT slope is 
$\Gamma \sim 1.95$--2.07 (with $K \sim 2.0$--$3.1 \times 10^{-2}$ photons keV$^{-1}$ cm$^{-2}$ s$^{-1}$; 
compare with Tables~\ref{t3} and \ref{t6}), yet some curvature is still present, as is a possible cutoff beyond 
$\sim 70$--80 keV (Fig.~\ref{dc}; see also Orr et al. 2001). An exponential break was then adopted, 
parameterized through the cutoff ($E_\rmn{C}$) and $e$-folding ($E_\rmn{F}$) energies.\footnote{The 
multiplicative high-energy cutoff \texttt{highecut} in \textsc{xspec} is defined as $\exp\,[(E_\rmn{C}-E)/E_\rmn{F}]$ 
for $E > E_\rmn{C}$, while the data below $E_\rmn{C}$ are not modified.}

\begin{table}
\caption{Key parameters of the extended fit to the 14--150 keV \textit{Swift}/BAT spectrum. $E_\rmn{C}$: 
exponential cutoff energy in keV. $E_\rmn{F}$: $e$-folding energy in keV. $\mathcal{C}$: cross normalization 
factor.}
\label{t7}
\begin{tabular}{cccc}
\hline
Model & \textsl{pcovpow} & \textsl{discref} & \textsl{pcovdbb} \\
\hline
$E_\rmn{C}$ & $> 70$ & 68$^{+21}_{-15}$ & 35$^{+22}_{-14}$ \\
$E_\rmn{F}$ & 100(f) & 100(f) & 98$^{+44}_{-51}$ \\
$\mathcal{C}$ & 1.22$\pm$0.06 & 0.58$\pm$0.02 & 0.90$\pm$0.05 \\
$\chi^2/\nu$ & 1921.6/1716 & 1903.3/1717 & 1812.1/1717 \\
\hline
\end{tabular}
\end{table}

The absorption and reflection frameworks are characterized by an opposite orientation towards the very hard 
X-ray spectral shape, although all the other quantities remain strictly consistent with the values obtained in the 
0.3--10 keV analysis, given the low statistical weight of the BAT spectrum. Specifically, the extrapolation of the 
intrinsic continuum with $\Gamma \sim 2.2$ in \textsl{pcovpow} falls short across $\sim 50$ keV, where the 
curvature simulates a sort of broad, residual hump that remains unfitted. The original goodness is preserved 
in \textsl{discref} through a standard rollover at $\sim 70$ keV, yet the most effective model is confirmed to be 
\textsl{pcovdbb} ($\Delta \chi^2/\Delta \nu \simeq 3.0/4$), with $E_\rmn{C} \sim 35$ keV and $E_\rmn{F} \sim 
100$ keV (Table~\ref{t7}). Also the cross normalization factor $\mathcal{C}$ between the EPIC/pn and BAT 
spectra is a valuable diagnostic. In the pure partial covering scenario we get $\mathcal{C} \sim 1.2$, meaning 
that during the \textit{XMM--Newton} observation the source has to be moderately underluminous with respect 
to its average state. In other words, the hard X-rays are somewhat underpredicted. The contrary occurs with 
disc reflection, for which $\mathcal{C} \sim 0.6$, while in the composite \textsl{pcovdbb} case the match is 
virtually optimal, with $\mathcal{C} \simeq 0.90(\pm 0.05)$. Such incongruences are potentially revealing, 
and are inherent to the models themselves. In fact, while any absorption effect rapidly decreases with energy, 
the reflection contribution grows, and the presence of a strong reflection component on top of a flatter $\Gamma 
\sim 1.8$ keeps the flux level above 10 keV significantly higher. The implications will be further discussed in the 
next Section. 

Switching to the low energy side, it is worth considering the source flux in the six \textit{XMM--Newton}/OM filters. 
It is quite striking that the extrapolation of the \textit{Swift}/BAT spectral shape generally overlaps with the optical 
points, with a slope just slightly steeper than $\Gamma \sim 2$. The variability of MR~2251$-$178 in the optical and 
X-ray bands is known to be very well correlated, with a mutual delay consistent with zero and anyway not larger than 
four days (Ar\'evalo et al. 2008), i.e. less than the span of this long-look observation. However, it is unlikely for a 
single continuum component to extend across nearly five decades in energy, with just the partial covering absorption 
effects producing an apparent broad trough at $\sim 1$--10 keV. Like the high-energy cutoff in the primary X-ray power 
law can be naturally related to the temperature of the relativistic electrons in the corona, a similar drop in response 
to the temperature of the seed photons from the disc must stand somewhere in the UV. With the central black hole 
accreting at a $\sim 0.1$--0.5 fraction of the Eddington rate, the inner disc temperature in MR~2251$-$178 is predicted 
to lie in the range from $\sim 10$ to 50 eV, irrespective of the exact spin (e.g. Peterson 1997). No substantial power-law 
contribution should then leak down to the OM bands. In any case, the optical and X-ray sources are tightly connected, 
and the former can be almost certainly identified with the accretion disc.

\begin{figure}
\includegraphics[width=8.5cm]{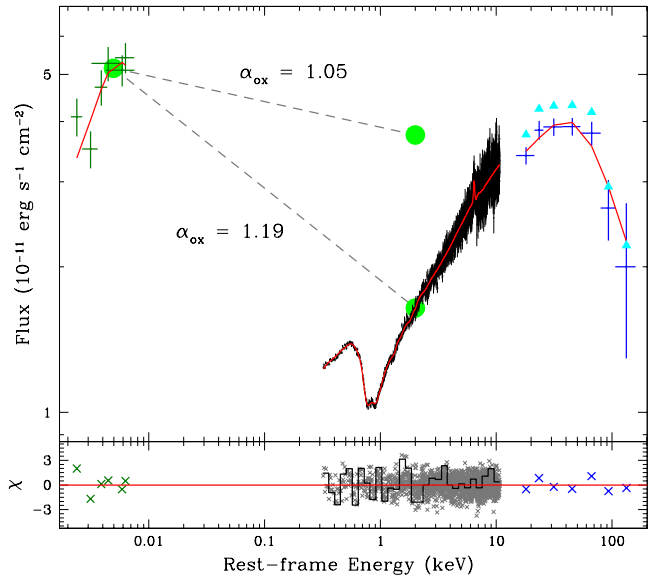}
\caption{Top panel: optical to hard X-ray spectral energy distribution of MR~2251$-$178, 
plotted in \textit{fluxed} units (i.e. $\nu f_\nu$) and obtained as a ratio over a power law with 
$\Gamma = 2$ (note that this is still an approximation; see the Appendix~A in Vaughan et al. 
2011 for a thorough discussion on the visualization of \textit{true} X-ray spectra). The red solid 
curve traces the best-fitting extended \textsl{pcovdbb} model, which includes a second, 
low-temperature disc component and a high-energy cutoff. The simultaneous \textit{XMM--Newton} 
OM and EPIC/pn data (corrected for Galactic extinction and absorption) are combined with the 
\textit{Swift}/BAT spectrum averaged over 70 months (blue crosses; the cyan points are shifted 
upwards by $\sim 10$ per cent according to the cross normalization factor, although it is most 
likely that the pair of \textit{XMM--Newton} spectra are slightly brighter than usual). The possible 
range for the optical to X-ray spectral index (no absorption correction and double partial covering 
case) is also shown. Bottom panel: residuals in units of $\sigma$. The error bars (of size 1) have 
been omitted.} 
\label{fs}
\end{figure}

The only model that intrinsically includes the disc thermal emission is \textsl{optxagn}. Neglecting the extremely poor 
fit statistics at 0.3--10 keV and any discrepancy with the BAT spectrum, once informed with the results of \textsl{pcovdbb} 
on e.g. $kT$ and $\Gamma$, \textsl{optxagn} yields a very smooth connection to the OM fluxes, with all the six 
points falling within 30 per cent of the extrapolation from the X-ray band for a range of sensible values of both the 
Eddington ratio and the coronal radius. It should be noted that departures from the adopted Galactic extinction law 
might be involved, as well as some extra reddening local to the source, possibly subject to the anomalous dust-to-gas 
ratios commonly found among AGN (e.g. Maiolino et al. 2001). Moreover, the OM fluxes could contain some contribution 
from the strong emission lines revealed by the \textit{HST} UV spectrum (Monier et al. 2001), in particular Mg~\textsc{ii} 
in UVW1 and C~\textsc{iii}] in UVW2. None the less, the observed correspondence suggests that the basic geometrical 
and physical assumptions underlying the \texttt{optxagnf} Comptonization picture (Done et al. 2012) are relevant to 
MR~2251$-$178, at least qualitatively. The unprocessed disc component has to be included in all our other models as 
a separate multi-temperature blackbody. Being constrained over the OM range only, the peak temperature is $kT \sim 
2$--3 eV, and the luminosity is $L_\rmn{disc} \sim 10^{45}$ erg s$^{-1}$. We consider this a conservative limit for the 
thermal emission from the outer disc. The optical to hard X-ray spectral energy distribution (SED) of MR~2251$-$178 is 
shown in Fig.~\ref{fs}, together with its best-fitting \textsl{pcovdbb} description.

%%%%%%%%%%%%%%%%%%%%%%%%%%%%%%%%%%%%%%%%%%%%%%%%%%%%%%%%%%%%%%%%%
\section{Discussion}

The 2011 \textit{XMM--Newton} observation supplied the highest-quality X-ray spectrum of MR~2251$-$178 
to date, with over 5 million net counts over the 0.3--10 keV band. This notwithstanding, based on the EPIC/pn 
analysis alone the origin of the intrinsic X-ray continuum would remain somewhat unclear, as partial covering 
absorption, ionized reflection, and Comptonized disc emission models cannot be statistically discriminated at 
CCD resolution. The availability of exquisite high-resolution data offers a straightforward way to test the 
soundness of the different fits, in which the column densities and ionization parameters of the warm absorption 
components have been allowed to vary with respect to the reference values of Table~\ref{t1}. Indeed, while 
various warm absorber configurations are still acceptable for the 0.3--10 keV continuum shape, it is evident 
from Fig.~\ref{nx} that in some cases the combined shifts in both $N_\rmn{H}$ and $\xi$ lead to a complete 
displacement of a given component within the parameter space. To a lesser extent, this is also true for the 
finest grid, which is the best-constrained in the broadband analysis due to its low ionization (e.g. Table~\ref{t3}). 
The quantity $\Delta\chi^2_{N_\rmn{H},\xi}$ has been used as a gauge of how much the warm absorption 
properties are distorted to minimize the fit statistics of each model. The deviations are largest in reflection-based 
models, which show strong residuals (mainly corresponding to the O~\textsc{vii} emission line) when superimposed 
on the RGS spectra.\footnote{We also emphasize that misread mid- and high-ionization zones would heavily 
affect the fit to the HETG spectra.} Not surprisingly, \textsl{pcovpow} is instead nearly consistent with R13, 
where an equivalent partially covered continuum was adopted. Yet, the closest agreement is delivered by 
\textsl{pcovdbb}, with $\Delta\chi^2_{N_\rmn{H},\xi} \sim 27$ (Table~\ref{t6}). Indeed, in trying to fit the 
EPIC/pn and RGS spectra simultaneously, \textsl{pcovpow} and \textsl{pcovdbb} turn out to be equally 
effective. 

As already mentioned, another useful diagnostic is the scale factor between the EPIC/pn and BAT 
spectra, especially because of the opposite trend entailed by the \textsl{pcovpow} and \textsl{discref} 
models. For ease of comparison, we have collected the about 40 measurements of the 2--10 keV flux 
obtained by ten X-ray observatories over more than three decades. The lowest and highest states 
correspond to the 1989 \textit{Ginga} observation (Mineo \& Stewart 1993) and to the first 1993 
\textit{ASCA} snapshot (Kaspi et al. 2004), respectively at $\sim 1.7$ and $5.2 \times 10^{-11}$ erg 
s$^{-1}$ cm$^{-2}$. The mean/median intensity is $\sim 3.2$--$3.3 \times 10^{-11}$ erg s$^{-1}$ 
cm$^{-2}$, suggesting that the 2011 \textit{XMM--Newton} campaign caught the source in a relatively 
bright phase, arguably $\sim 25$--30 per cent above the average. Qualitatively, this would disfavour 
the pure partial covering picture, but it is not yet a conclusive hint in any direction for several reasons. 
The magnitude of the cross normalization actually casts serious doubts also on \textsl{discref}, whose 
application to the 2009 \textit{Suzaku} data set confirms that the sizable offset between the 0.6--10 
and 15--50 keV spectra is not related to different flux states. The small value of $\mathcal{C} \sim 
0.6$ and the cutoff at $\sim 70$ keV would then compensate for the obvious absence of a prominent 
Compton hump, which is perhaps related also to the clear pattern in the fit residuals starting above 
9 keV (Fig.~\ref{dm}).

On the other hand, cold absorption effects can easily halve the 2--10 keV emission, artificially 
expanding the scope of the intrinsic X-ray variability;\footnote{Adopting a conservative binning of 
$\sim 3$--4 months to obtain similar error bars on each point, the 70-month \textit{Swift}/BAT light 
curve shows peak-to-peak variations by a factor of $\sim 2.5$.} the source might simply look fainter 
because it is more obscured. In this view, even a cross normalization slightly larger than one could 
be broadly consistent with the observational history 
of MR~2251$-$178, also given that $\mathcal{C}$ strongly depends on the actual photon index. 
Moreover, in the regular monitoring by \textit{RXTE} discussed in Arevalo et al. (2008), consisting 
of over 200 snapshots in 2.5 years with 1 ks exposure each, the inferred 2--10 keV flux covers the 
$\sim 2.8$--$7.3 \times 10^{-11}$ erg s$^{-1}$ cm$^{-2}$ range. Disregarding any possible systematics 
in the latter estimates, their average value of $5 \times 10^{-11}$ erg s$^{-1}$ cm$^{-2}$ is about 
20 per cent larger than the flux recorded in the \textit{XMM--Newton} observation, in agreement 
with the extrapolation to the \textit{Swift}/BAT spectrum of \textsl{pcovpow}, which anyway returns 
a very poor fit around 50 keV. As a consequence, the composite model \textsl{pcovdbb} is definitely 
preferred. An accurate, simultaneous broadband analysis, which has now become feasible at $\sim 
3$--80 keV thanks to \textit{NuSTAR} (Harrison et al. 2013), is clearly needed to shed new light on the 
nature of the high-energy emission of MR~2251$-$178, although the soft X-rays cannot be neglected 
to properly constrain the warm absorber and the putative soft excess. 

The global curvature is undoubtedly the most puzzling feature of the broadband X-ray continuum in 
this source, with its strikingly different slopes below and above 10 keV. A similar behaviour has 
been observed in a completely different object, the dwarf Seyfert galaxy NGC 4395, known to host 
the least massive black hole among AGN ($M_\rmn{BH} \sim 10^5 M_{\sun}$; Peterson et al. 2005). 
Its 2--10 keV spectral shape is extraordinarily flat, characterized by a photon index that can be as low 
as $\sim 0.6$ (Moran et al. 2005), yet turning into the typical $\Gamma \sim 2$ of unobscured AGN at 
higher energies. NGC 4395 actually undergoes dramatic spectral variability over timescales of a few 
ks, which can be ascribed to a complex system of cold absorbers with column densities of $\sim 
10^{22}$--10$^{23}$ cm$^{-2}$, crossing the line of sight to the primary source (Nardini \& Risaliti 2011). 
It is then possible that in MR~2251$-$178 the observed 2--10 keV continuum is misleading, not being 
representative of the intrinsic emission without a correct modelling of the absorption effects. Indeed, 
such a hard slope poses an issue with the depth of the inner-shell lines in the low-ionization warm 
absorber, which call for a soft $\Gamma \sim 2.5$ illumination. Overall, the grating analysis suggests 
a break in the input photoionizing spectrum, with a sort of intrinsic soft excess converting into a much 
flatter power law than typically found in radio-quiet quasars. 

Following R13, who suggest an association with the same BLR clouds responsible for the soft X-ray 
emission lines, we assume that the partial-covering gas is more internal than the pc-scale warm absorber, 
and approximate the radiation impinging on the latter as a broken power law in order to assess its crude 
outline. Within \textsl{pcovpow}, the soft photon index is bound to the $\Gamma \sim 2.2$ of the unabsorbed 
continuum, which is possibly not steep enough to comply with the prescriptions from the warm absorption 
lines. The extra blackbody-like component included in the \textsl{pcovdbb} model results in a broadly 
acceptable effective slope of $\Gamma \sim 2.3$ below the break energy of $\sim 1.2$ keV. The implied 
temperature of $\sim 0.14$ keV is at least three times larger than that expected in MR~2251$-$178 for a 
standard thin accretion disc, but it is commensurate with the typical values recorded for the soft X-ray 
excess in AGN (e.g. Jin et al. 2012), whose origin is still essentially unknown. Despite these uncertainties, 
the combination of cold absorption in the form of partial covering and some kind of Compton reprocessing 
stands as the most convincing interpretation on the whole, also conceding that the Comptonization conjecture 
lends weight to the tight connection between the accretion disc and the X-ray corona (Arevalo et al. 2008). 

Given that the X-ray source in AGN is presumed to be extremely compact (a few tens of gravitational 
radii at most), partial covering can be only envisaged in a dynamical context; in other words, it has to 
occur at BLR scales. It is easy to show, in fact, that if no changes in the covering factor are perceived 
over several years the putative absorber must be located at least a few pc away from the emitting 
region, and be much larger of it. Under these circumstances, any partial covering configuration is highly 
unlikely. Concerning MR~2251$-$178, the lack of any appreciable changes other than the slow decline 
in the overall flux intensity during the \textit{XMM--Newton} long look is not particularly worrying given 
the large mass of the central black hole. The relevant timescale to the variations of the intrinsic X-ray 
emission, for instance, is the light-crossing time, which is $\sim 12$ ks assuming that the source is 
spherical and has a diameter of $10\,r_\rmn{g}$ (compare with Fig.~\ref{lc}). On the other hand, within 
an eclipsing cloud scenario (e.g. NGC~1365; Risaliti et al. 2007), the occultation time by a single gaseous 
clump with identical size and shape to the X-ray source in Keplerian motion at a distance of $10^4\,r_\rmn{g}$ 
would be nearly 2 weeks,\footnote{It is 30 minutes in NGC~4395 under the same assumptions.} against the 
5.5 days of elapsed time in present observation. By using the recent $R_\rmn{BLR}$--$\lambda 
L_\lambda$(5100\,\AA) relationship from Bentz et al. (2013) and the average 
5100\,\AA~luminosity from Lira et al. (2011), we indeed obtain a BLR radius of about 75 light days, i.e. 
$5.5 \times 10^3\,r_\rmn{g}$. Notably, as also reported in R13, there are hints of some change in the 
covering fractions between the \textit{XMM--Newton} and \textit{Chandra} campaigns, separated by just 
a few weeks. 

The partial covering interpretation for the X-ray spectra of AGN thus relies on the identification 
of rapid spectral variability. Keeping in mind the expanded timescale, the long look of MR~2251$-$178 
is equivalent to the shorter snapshots of NGC 4395, which sampled a single state of the source. 
In this framework, we can draw similar speculations on the structure of the cold X-ray absorber, 
holding for both \textsl{pcovpow} and \textsl{pcovdbb}. Two distinct components emerge from the 
pure partial covering fit. In a stratified circumnuclear medium, the possibility that two \textit{clouds} 
belonging to detached layers simultaneously cover different sectors of the X-ray source is rather 
contrived. In principle, the disparity of about one order of magnitude between the column densities 
involved would allow for projection effects, with the two components partially on top of each other. 
The assessed covering fraction $f_1 \simeq 0.23$ would be simply a lower limit to the effective one. 
Yet the fact that both $N_\rmn{H,1}$ and $N_\rmn{H,2}$ are very well constrained, and in particular 
that the uncertainty on $N_\rmn{H,2}$ is only $\sim 0.5\,N_\rmn{H,1}$ (at the 90 per cent confidence 
level), seems to rule out this solution. On the contrary, a single absorber can meet the fit conditions 
if bearing sharp density gradients and/or irregular geometry. Indeed, it has been recently suggested 
that the eclipsing BLR clouds might actually have a comet-like structure, with a dense head and a 
tenuous tail (Maiolino et al. 2010; Risaliti et al. 2011). Such blobs could have a lifetime of several 
months at most, implying that this BLR component must be continuously replenished. It is therefore 
a matter of duty cycle and likelihood for a gas clump to pass in front of the X-ray source. The transit 
of a single cloud, further supported by the robustness of the \textsl{pcovdbb} picture, is ultimately 
the most reliable scenario in this sense. Limited to the column density changes of the warm absorber, 
the presence of material moving in and out the line of sight had already been invoked (e.g. Orr et al. 
2001; Kaspi et al. 2004). This bulk motion might exist throughout the spatial scales. The spectral 
variability aspects are then crucial for any of the physical models examined here, and will be 
investigated in detail in a subsequent paper (Porquet et al., in preparation) considering all the 
highest-quality recent and archival observations of MR~2251$-$178. 

Summarizing, our analysis still leaves some open issues. However, irrespective of the real nature of its X-ray 
emission and of the amount of absorbing gas along the line of sight, MR~2251$-$178 is confirmed to boast 
one of the largest X-ray to optical luminosity ratios among radio-quiet quasars. This is clearly pointed out by 
its $\alpha_\rmn{ox}$ spectral index.\footnote{The optical/UV to X-ray spectral index is generally defined as 
$\alpha_\rmn{ox} = -0.384 \log\,(L_\rmn{2\,keV}/L_\rmn{2500\,\text{\AA}})$. Note that $\alpha_\rmn{ox} = 1$ 
for a flat SED (in $\nu f_\nu$).} Without applying any correction, the observed $\alpha_\rmn{ox} \simeq 1.19$ 
extracted from the broadband SED of Fig.~\ref{fs} is already much lower than the mean value for X-ray selected 
AGN, i.e. $\overline{\alpha_\rmn{ox}} \simeq 1.37$ (with a 1$\sigma$ dispersion of 0.18; Lusso et al. 2010). 
In the dual partial covering case, the rest-frame brightness at 2 keV has to be revised upwards by a factor of 
$\sim 2.3$, delivering an even more exceptional optical to X-ray spectral index of 1.05. This does not heavily 
impact the bolometric output of MR~2251$-$178 though, once the correlation between $\alpha_\rmn{ox}$ 
and the 2--10 keV to bolometric correction ($k_\rmn{bol}$) is taken into account. We obtain that $k_\rmn{bol} 
\approx 7$--12 (e.g. Lusso et al. 2010; Marchese et al. 2012), for a total luminosity $L_\rmn{bol}$ of $\sim 
5$--7$\times$10$^{45}$ erg s$^{-1}$, which is $\sim 15$--25 per cent of the source Eddington luminosity 
and is just above the UV-based estimate (Dunn et al. 2008). In the light of these outstanding properties, 
including the possible massive disc outflow (Gibson et al. 2005; Gofford et al. 2011), unveiling the X-ray 
emission mechanism and its coupling with the absorbing/reprocessing gas would have far-reaching 
implications for the quasar radiative and mechanical feedback on the surrounding environment and 
all the related fields.
 
%%%%%%%%%%%%%%%%%%%%%%%%%%%%%%%%%%%%%%%%%%%%%%%%%%%%%%%%%%%%%%%%%
\section{Summary and Conclusions}

Based on a recent \textit{XMM--Newton} long-look observation with $\sim 270$ ks of net exposure, 
we have reported on the 0.3--10 keV EPIC/pn spectral analysis of MR~2251$-$178, one of the brightest 
radio-quiet quasars in the local Universe. Following a companion paper dedicated to the study of the 
high-resolution grating spectra and of the physical properties of the complex warm absorber, here we 
have focused on the nature of the intrinsic X-ray emission. The broadband X-ray continuum of 
MR~2251$-$178 is known to exhibit substantial curvature up to $\sim 100$ keV, where a possible cutoff 
is also present. Together with the apparent soft excess below 0.7 keV and the warm absorption trough at 
$\sim 1$--2 keV, this gives rise to a peculiar steep--flat--steep spectral shape, which has been inspected 
within the frameworks of partial covering absorption, ionized reflection, and Comptonized disc emission. 

As their application turns out to be nearly statistically equivalent over the 0.3--10 keV band, all these 
models have also been compared with the coeval OM photometric data and the $\sim 14$--150 keV 
\textit{Swift}/BAT spectrum, averaged over 70 months. Assuming no cross normalization, the entire SED 
shows that the optical and hard X-ray fluxes lie at almost the same level in $\nu f_\nu$ units. Indeed, the 
optical to X-ray spectral index is extremely small ($\alpha_\rmn{ox} < 1.2$), irrespective of the exact 
absorption correction. The corresponding bolometric output of MR~2251$-$178 is estimated to be 
$L_\rmn{bol} \sim 5$--$7 \times 10^{45}$ erg s$^{-1}$, i.e. $\sim 15$--25 per cent of the source Eddington 
luminosity. 

None of the alternative interpretations taken into account in this work are conclusive by themselves, 
displaying both strengths and limitations at the same time, which can be summarized as follows:

1) The pure partial covering scenario reveals two distinct low-ionization components with columns of 0.6 and 
$6.6 \times 10^{23}$ cm$^{-2}$, eclipsing a fraction of $\sim 0.23$ and 0.49 of the X-ray source, respectively. 
We argue that these absorbers are located at BLR scales, but are unlikely to belong to different layers. Density 
gradients might be entailed, pointing to a single clump of gas with a dense core and a lighter halo. The intrinsic 
photon index $\Gamma \sim 2.2$ is close to what is required by the photoionization models of the warm absorber, 
but still somewhat lower; even so, excess curvature is left at hard X-rays around 50 keV. 

2) At EPIC/pn resolution, X-ray reflection provides a reasonable fit without implying any extreme gravity regime. 
The blurred disc component, however, is unusually cold, and $\sim 50$ times less ionized than its distant 
counterpart. This is possibly an artefact to compensate for the inadequate results with the soft X-ray emission 
lines (resolved by the gratings), and also leads to a misrepresentation of the warm absorption properties (for 
which the illumination is anyway flat). The degree of smoothing appears too large for the iron K band, and a clear 
pattern is found in the residuals above 9 keV. Moreover, the BAT spectrum is largely overpredicted. 

3) The optically-thick/thin Comptonization of thermal photons from the inner disc into both the soft excess and 
the hard power law, with the outer regions emitting as a colour temperature corrected blackbody, is established 
as the most obvious connection between the optical and X-ray bands. Nevertheless, in spite of the remarkably 
good extrapolation to the OM data and of the ample leverage on the physical parameters, a basic self-consistent 
model either fails to reproduce the 0.3--10 keV features or invariably plummets too quickly at higher energies. 
Any agreement is lost on the three warm absorption components. 

A composite model allows us to successfully overcome most of the above shortcomings. While reflection from 
the accretion flow and/or the ambient material might still be involved to a certain extent, in the preferred picture 
the hard power-law continuum ($\Gamma \sim 1.75$) steepens below $\sim 1$ keV into a soft excess with 
$kT \sim 0.14$ keV, whose origin remains unclear although it can be tentatively associated with Comptonization 
in the disc atmosphere. A single cloud with $N_\rmn{H} \sim 3 \times 10^{23}$ cm$^{-2}$ is present along the line 
of sight, covering about one sixth of the X-ray source. 

In conclusion, the X-ray observations of MR~2251$-$178 brought to light a complex and stratified 
environment close to the central source, which deeply transforms the shape of the intrinsic X-ray 
spectrum. Our study suggests that only a combination of time variability analysis, simultaneous 
broadband coverage and high spectral resolution can help disentangling the effects of the different 
physical processes responsible for the observed X-ray emission of AGN. 

%%%%%%%%%%%%%%%%%%%%%%%%%%%%%%%%%%%%%%%%%%%%%%%%%%%%%%%%%%%%%%%%%%%%
\section*{Acknowledgments}

The authors are grateful to the referee, Chris Done, for helpful comments that significantly improved the clarity 
of this paper. EN thanks Javier Garc\'ia for useful discussion on the reflection strength in his \texttt{xillver} table 
models. EN, JNR and JG acknowledge the financial support provided by STFC. JNR was also supported by 
\textit{Chandra} grant number GO1-12143X. DP acknowledges financial support from the French GDR PCHE.
This work is based on observations obtained with \textit{XMM--Newton}, an ESA science mission with 
instruments and contributions directly funded by ESA member states and NASA. 

%%%%%%%%%%%%%%%%%%%%%%%%%%%%%%%%%%%%%%%%%%%%%%%%%%%%%%%%%%%%%%%%%%%%%%%%%%%

%%%%%%%%%%%%%%%

\label{lastpage}

\end{document}